\documentstyle[psfig,11pt]{article}
\def\OO#1{{\cal O}(c^{-#1})}
\def\ve#1{{\bf #1}}
\def\vecg#1{\mbox{\boldmath$#1$}}
\def\ov#1{{\overline{#1}}}

\def\TCG{{\sf TCG}}
\def\TCB{{\sf TCB}}
\def\TT{{\sf TT}}
\def\TDB{{\sf TDB}}

\newcommand{\muas}[0]{\hbox{\rm $\mu$as}}

\newcommand{\arcsec}[0]{\hbox{$^{\prime\prime}$}}

\def \noi {\noindent}

\def \ssk{\smallskip}

\begin{document}

\vspace*{1.2cm}

\noi {\Large
Earth's rotation in the framework of general relativity:
rigid multipole moments
}

\vspace*{1cm}
\noi \hspace*{1.5cm} S.A. Klioner, M. Soffel, Ch.Xu, X.Wu\\
\noi \hspace*{1.5cm} Lohrmann Observatory, Dresden Technical University\\
\noi \hspace*{1.5cm} 01062 Dresden, Germany\\

\vspace*{2cm}

\noi {\large ABSTRACT.} A set of equations describing the rotational
motion of the Earth relative to the GCRS is formulated in the
approximation of rigidly rotating multipoles. The external bodies are
supposed to be mass monopoles. The derived set of formulas is supposed to
form the theoretical basis for a practical post-Newtonian theory of Earth
precession and nutation.

\vspace*{1cm}

\noi {\large 1. Introduction}

\vspace*{5mm}

The relation between the International Celestial Reference System
(ICRS) and the corresponding terrestrial one (ITRF) is a central
problem of astrometry, geodesy and related disciplines. From a
theoretical point of view this requires not only a precise
determination of the ICRS and the ITRS, but also a detailed modelling
of Earth's rotation. Due to the high accuracy requirements it is
obvious that all of these problems have to be formulated in Einstein's
theory of gravity, at least in its first post-Newtonian approximation.
Until now most Newtonian treatments of Earth's rotation are based upon
some highly accurate rigid body theory such as SMART97 (Bretagnon {\it
et al.}, 1998) and add effects from elasticity, the atmosphere, the
oceans, the core etc. in a perturbative manner. Actually, the concept
of a rigid body is very powerful in Newton's theory where the three
fundamental axes, the total angular momentum or spin axis, the rotation
axis and the figure axis, can be introduced without efforts.
Unfortunately rigid bodies with an internal velocity field of the form
$\ve{v}=\vecg{\omega}\times\ve{x}$ in general do not exist in General
Relativity. Nevertheless one might introduce a certain class of models,
where the time behavior of potential coefficients, moments of inertia
tensor, etc. is completely determined by some quantity
$\vecg{\omega}(T)$. We call such models ``rigidly rotating multipole''
models.

The aim of this paper is to summarize a set of formulas describing the
rotational motion of the Earth with respect to the Geocentric Celestial
Reference System (GCRS) in the approximation of rigidly rotating
multipole moments. The GCRS is defined in the post-Newtonian approximation of
General Relativity by the IAU Resolution B1.3 adopted at the 24th
General Assembly of the IAU (Manchester, 2000) and published in the IAU
Information Bulletin No. 88 (see also erratum in Bulletin No. 89). Full
text of this Resolution can be also found at {\tt
http://danof.obspm.fr/IAU\underline{\ }resolutions/Resol-UAI.htm}. The approximation
of rigidly rotating multipoles used in this paper is a phenomenological
model which allows one to simplify the mathematical description of
rotational motion almost to the level of Newtonian theory.
Likely, the model of rigidly rotating multipoles is not consistent with
general relativity in the sense that there is no physical equation of state and local
conditions of matter that will support the model.
However, a theory of motion
within this model can be used as a first approximation to be refined
later by methods of perturbation theory. Within this model the accuracy
of the given formulas is not worse than 0.1 \muas.

\vspace*{1cm}

\noi {\large 2. Post-Newtonian equations of rotational motion}

\vspace*{5mm}

The post-Newtonian equations of rotational motion of the Earth relative
to the Geocentric Celestial Reference System (GCRS) can be derived from
the metric tensor of the GCRS (Voinov, 1988; Damour, Soffel, Xu, 1993;
Klioner, 1996) in the form

\begin{equation}\label{eqrm}
{d\over d\,TCG}\, S^a=L^a+\OO4.
\end{equation}

\noindent
where $S^a$ is the post-Newtonian spin defined as an explicit integral
over the body of the Earth

\begin{eqnarray}\label{pN:spin}
S^a&=&\varepsilon_{abc} \int_{V} X^b\, p^c\, d^3X
+\OO4.
\end{eqnarray}

\noindent
Here,

\begin{eqnarray}\label{pN:Q}
p^a&=&\Sigma^a\, \left(1+ {4\over c^2}\, W\right)
-{1\over 2c^2}\,G\,\Sigma
\int_{V} \Sigma^b(T,\ve{X}')\,\,{7\,\delta^{ab} +
n^a n^b \over |\ve{X}-\ve{X}'|}\,\, d^3X'+{\cal O}(c^{-4}),
\end{eqnarray}

\begin{eqnarray}\label{ni}
n^a&=&{X^a-X'^a\over |\ve{X}-\ve{X}'|},
\end{eqnarray}

\noindent
and $\Sigma$ and $\Sigma^a$ are defined by the components of the
energy-momentum tensor in the GCRS, ${\cal T}^{\alpha\,\beta}$, as
$\Sigma={\cal T}^{\alpha\,\alpha}$ and $\Sigma^a={\cal T}^{0a}$.
$W$ is the potential appearing in the metric tensor of the GCRS. This
definition of spin was first derived by Fock (1955) and thoroughly
discussed in, e.g., Damour, Soffel and Xu (1993). The right-hand side
of (\ref{eqrm}) represent the post-Newtonian torque which can be
represented as

\begin{eqnarray}\label{L-BD}
L^a&=&
\sum_{l=0}^\infty\ {1 \over l!}\ \left(
\varepsilon_{abc}\, M_{bL}\, G_{cL} + {1 \over c^2}\, {l+1 \over l+2}\
\varepsilon_{abc}\, S_{bL}\, H_{cL} \right)
%???
+{d\over d\,\TCG}\widetilde S^a.
%???
\end{eqnarray}

\noindent
Here $M_L$ and $S_L$ are the Blanchet-Damour mass and spin multipole
moments characterizing the Earth, and $G_L$ and $H_L$ for $l\ge 2$ are
the gravitoelectric and gravitomagnetic tidal moments of the external
gravitational field experienced by the Earth. The moments $M_L$ are
equivalent to the set of coefficients $C_{lm}$ and $S_{lm}$ in the
conventional expansion of the gravitational potential of the Earth in
terms of spherical functions. Explicit formulas for $G_L$ within the
adopted model will be given below. The moment $H_a$ describes the
inertial forces induced by the rotation of the GCRS relative to the
locally inertial reference system (these forces appear because the GCRS
is defined to be kinematically non-rotating with respect to the BCRS):

\begin{equation}
\label{Omega-precession}
\Omega^a_{\rm iner}={1\over 2c^2}\,H_a=
-{3\over 2\,c^2}\,\varepsilon_{aij}\,
v_E^i\,{\partial\over \partial x^j} w^{\rm ext}(\ve{x}_E)
+{2\over c^2}\,\varepsilon_{aij}\,
{\partial\over \partial x^j} w^i_{\rm ext}(\ve{x}_E)
-{1\over 2\,c^2}\,\varepsilon_{aib}\,
v_E^i\,G_b,
\end{equation}

\noindent
where $v_E^i$ is the barycentric velocity of the Earth, $w^{\rm
ext}(\ve{x}_E)$ and $w^i_{\rm ext}(\ve{x}_E)$ are the external BCRS
potential evaluated at the geocenter, and $G_a$ is the acceleration of
the geocenter with respect to the geodetic motion (all these quantities
are defined and briefly explained in the IAU Resolution B1.3). The quantity
$\Omega^a_{\rm iner}$ represent the angular velocity of precession of the GCRS
with respect to the locally inertial axis. This relativistic precession
consist of geodetic, Lense-Thirring and Thomas precessions (the three
terms in (\ref{Omega-precession}), respectively) and amounts to $\sim
1.92\arcsec$ per century plus periodic terms with maximal amplitudes of
0.15 mas. It is easy to estimate that
$|\Omega^a|<3.1\cdot10^{-15}\ {\rm s}^{-1}$.

%???
The last term in (\ref{L-BD}) represent a total time derivative and can
be in principle included in the definition of the post-Newtonian spin
$S^a$ as suggested by Damour, Soffel and Xu (1993). Here we prefer to
retain the definition of $S^a$ to be (\ref{pN:spin})--(\ref{ni}). An
explicit formula for $\widetilde S^a$ is given by Eqs. (2.14)--(2.16)
of Damour, Soffel, Xu (1993). Numerical estimates of this term and the
way to cope with it will be published elsewhere.
%???

\vspace*{1cm}

\noi {\large 3. Post-Newtonian angular velocity and tensor of inertia}

\vspace*{5mm}

In order to be able to discuss the rotational motion of the Earth it is
not sufficient to consider only the time dependence of the spin $S^a$
described by (\ref{eqrm}). In classical Newtonian (Eulerian) theory of
a gyroscope the concepts of a figure axis, tensor of inertia and an
angular velocity of the body play a central role and one expects the
introduction of corresponding quantities in a relativistic framework to
be very fruitful.

Different approaches leading to the same results are possible.
Both restricted rigid body models (Thorne, G\"ursel, 1983; Soffel, 1994)
and a theory of post-Newtonian Tisserand axes for a deformable
body (Klioner, 1996) allows one to derive the same definition
of the post-Newtonian tensor of inertia and split
the post-Newtonian spin $S^a$ defined by (\ref{pN:spin})--(\ref{ni})
in the Newtonian-looking way

\begin{equation}
\label{S-C-omega}
S^a=C^{ab}\,\omega^b,
\end{equation}

\noindent
where $C^{ab}$ is the post-Newtonian tensor of inertia
and $\omega^b$ is the
angular velocity of rotation of the post-Newtonian Tisserand axes
(Klioner, 1996).

The explicit formula for $C^{ab}$ as an integral over the volume of the
Earth is given in Klioner (1996).
Although the definition of
$C^{ab}$ contains a number of explicit relativistic terms usually we do
not compute $C^{ab}$ from the distribution of density, pressure, etc.
within the Earth, but determine the values of $C^{ab}$ from
observations (as numerical parameters of the models). Therefore, for
practical purposes we can simply use the
fact that the spin $S^a$ can be represented in the form given in
(\ref{S-C-omega}).

\vspace*{1cm}

\noi {\large 4. Rigidly rotating multipole moments}

\vspace*{5mm}

Up to now the tensor of inertia $C^{ab}$ and the multipole moments of
the Earth's gravitational field $M_L$ and $S_L$ were considered as
arbitrary functions
of time. In Newtonian theory of Earth's rotation a rigid Earth plays a
very important role as a first order approximation. This rigid model i)
crucially simplifies the mathematical description of the rotational
motion and ii) is not too far from reality, so that the effect of
non-rigidity can be than added to the model by means of perturbation
theory. The reason why the rigid body model substantially simplifies
the rotational equations of motion is that both the mass multipole
moments $M^{\rm Newt}_L$ and the tensor of inertia $C^{ab}_{\rm Newt}$
rotate rigidly with the same angular velocity. In other words there
exist a rigidly rotating reference system $Y^a=P^{ab}(T)\,X^a$, where
$P^{ab}(T)$ is some time-dependent orthogonal matrix, where both
$M^{\rm Newt}_L$ and $C^{ab}_{\rm Newt}$ are constant. Moreover, in
Newtonian theory one can easily prove that the time dependence of
matrix $P^{ab}(T)$ defines the same angular velocity which appears in
the Newtonian analog of (\ref{S-C-omega}). All this can be proved
starting from the fundamental assumption that the velocity $\ve{v}$ of
matter inside the body is described by a rigid rotation
$\ve{v}=\vecg{\omega}\times\ve{X}$.

It is well known that in general relativity it is impossible to define
a rigid non-isolated body in a self-consistent way even in the first
post-Newtonian approximation (see, e.g., Thorne, G\"ursel, 1983).
However, we can {\sl assume} the same
nice properties of the relativistic tensor of inertia $C^{ab}$ and
multipole moments $M_L$ and $S_L$ that we had in Newtonian theory.
Thus, we define the model of rigidly rotating multipole moments
by means of a series of {\sl assumptions}:

\begin{eqnarray}\label{ovCij}
&&C^{ab}=P^{ac}\,P^{bd}\,\ov{C}^{cd},
\quad \ov{C}^{cd}={\rm const}
\\
\label{ovM}
&&M_{a_1a_2\dots a_l}=P^{a_1b_1}\,P^{a_2b_2}\dots P^{a_lb_l}\,
\ov{M}_{b_1b_2\dots b_l},\quad
\ov{M}_{b_1b_2\dots b_l}={\rm const},
\quad l\ge2,
\\ \label{SL-CiL-omegai}
&&S_{L}=C^{bL}\,\omega^b,\quad l\ge2,
\\ \label{CiL-rot}
&&C^{ba_1a_2\dots a_l}=\,
P^{bd}\,P^{a_1c_1}\,P^{a_2c_2}\dots P^{a_lc_l}\,
\ov{C}_{dc_1c_2\dots c_l},\quad
\ov{C}_{dc_1c_2\dots c_l}={\rm const},
\quad l\ge 2,
\end{eqnarray}

\noindent
where $P^{ab}(T)$ is the orthogonal matrix  related to $\omega^a$ by
the kinematical Euler equations

\begin{eqnarray}\label{omegai-Pij}
\omega^a(T)={1\over2}\,\varepsilon_{abs}\,P^{db}(T)\,{d\over d\TCG}\,P^{dc}(T).
\end{eqnarray}

\noindent
Note that we {\sl assume} that $\omega^a$ in (\ref{SL-CiL-omegai}) and
(\ref{omegai-Pij}) is identical with $\omega^a$ from (\ref{S-C-omega}).
We {\sl assume} also that to bring $\ov{C}^{ab}$ into a diagonal form
one more {\it time-independent} rotation is required. The matrix
$P^{ab}(T)$ can be parametrized by three Euler angles $\psi$, $\theta$,
$\varphi$, and the time-derivatives of these angles define the angular
velocity of rotation according to (\ref{omegai-Pij}). Relations
(\ref{SL-CiL-omegai})--(\ref{CiL-rot}) for the higher spin moments
$S_L$, $l\ge 2$ and for $C_{iL}$ are only necessary to Newtonian
accuracy since they appear only in relativistic terms of (\ref{L-BD}).

Let us stress again that in Newtonian theory one one can derive
(\ref{S-C-omega}) and (\ref{ovCij})--(\ref{omegai-Pij}) from the
fundamental property of rigidity of the body
$\ve{v}=\vecg{\omega}\times\ve{X}$. On the contrary, in general
relativity we define the model by {\sl assuming} the properties of
$C^{ab}$, $M_L$ and $S_L$ without further restrictions of the local
flow of matter.

The experience of Newtonian models of Earth's rotation shows that the
phenomenological model (\ref{ovCij})--(\ref{omegai-Pij}) can be used as
a first-order approximation for a description of the global rotational
motion of the Earth. As in Newtonian theory, such a model serves as a
basis for considering the effects of non-rigidity in the rotational motion
of the Earth.

\vspace*{1cm}

\noi {\large 5. Further simplifying assumptions}

\vspace*{5mm}

A number of additional simplifying assumptions will be adopted here.
Some of these assumptions are justified by numerical estimations of the
corresponding terms in the equations of rotational motion of the Earth.

\begin{itemize}

\item
It is natural to assume that the GCRS is defined in such a way that the
mass dipole $M_a$ of the Earth vanishes (this assumption is actually
supported by the IAU Resolution B1.4 in (IAU, 2001)), i.e. the origin of the GCRS
is assumed to agree with the post-Newtonian center of mass of the Earth.

\item
In order to numerically estimate the influence of the terms in the
right-hand side of (\ref{L-BD}) produced by $S_L$ with $l\ge2$ (and for
that purpose only!) let us consider the following model for the matter of
the Earth

\begin{equation}\label{Sigma-a=rigid}
\Sigma^a=\Sigma\,\varepsilon_{abc}\,\omega^b\,X^c.
\end{equation}

\noindent
Since we want to numerically estimate post-Newtonian terms it is sufficient
to consider (\ref{Sigma-a=rigid}) as a Newtonian assumption
of a rigidly rotating body. Substituting (\ref{Sigma-a=rigid})
into the definition of $S_L$ for $l\ge2$ one can prove that

\begin{equation}\label{S-C-omega-Newtonian}
S_L = C_{aL}\,\omega^a,
\end{equation}

\begin{equation}\label{C-M-N}
C_{aL} = - M_{aL}+{l+1\over 2l+1} \delta_{a<b_l} N_{L-1>},
\end{equation}

\noindent
where the moments $M_L$ and $N_L$ to Newtonian order read
\begin{eqnarray}\label{M-N}
M_L &\equiv& \int_E \Sigma\, \hat X^L \, d^3X, \\
N_L &\equiv&  \int_E \Sigma\, \ve{X}^2 \,\hat X^L \, d^3X.
\end{eqnarray}

\noindent
Here the angle brackets ``$<\dots>$'' as well as the caret
``$\,\,\hat{}\,\,$'' indicate symmetric and tracefree (STF)
part of the corresponding expression. Eqs.
(\ref{S-C-omega-Newtonian})--(\ref{M-N}) allow one to estimate the
torque $\left.\delta\dot S^a\right|_{S_L}$ due to $S_L$, $l\ge2$ as

\begin{eqnarray}
\left.\left(|\delta\dot S^a|\over |S^a|\right)\right|_{S_L}\sim
\,\max(|J^E_{l+1}|,|J^E_{l-1}|)\,\cdot\,\sum_A {GM_{_A}\over c^2}\,{v_{EA}\over r_{EA}^2}\,
\left({R_E\over r_{EA}}\right)^{l-1}<10^{-20}\,{\rm s}^{-1}.
\end{eqnarray}

\noindent
This estimate gives typical angular velocity of precession due to
$S_L$. This can be compared, e.g., with the relativistic precession due
to $\vecg{\Omega}_{\rm iner}$. It is easy to see that the precession
due to $S_L$ for $l\ge2$ is at least a factor $10^{5}$ smaller than the
relativistic precession due to $\vecg{\Omega}_{\rm iner}$ (i.e. than
the geodetic precession). This implies that these terms for any $l\ge
2$ can be neglected at the accuracy level of 0.1 \muas. This
circumstance makes the assumption
(\ref{SL-CiL-omegai})--(\ref{CiL-rot}) for $S_L$, $l\ge2$ superfluous.

\item
All external bodies are supposed to be mass-monopoles, that is point
masses characterized only by their masses $M^A$ and BCRS positions
$\ve{x}_A(t)$, $t=\TCB$. Since the multipole structure (e.g.,
oblateness) of external bodies is not taken into account in the modern
Newtonian theories of nutation of the rigid Earth, such an assumption
does not prevent us to achieve the required accuracy of 0.1 \muas\ also
in the relativistic framework. In the framework of this model one can
derive explicit formulas for the external tidal moments $G_L$
influencing the Earth ($E$). One has

\begin{eqnarray}\label{Q-q}
G_{L}=\sum_{A\neq E}\,GM_A\,g_{L}^A,
\end{eqnarray}

\noindent
where $g_{L}^A$ are functions of i) the BCRS position $\ve{x}_{E}$,
velocity $\ve{v}_{E}$ and acceleration $\ve{a}_{E}$ of the Earth, ii)
the BCRS position $\ve{x}_{A}$, velocity $\ve{v}_{A}$ and acceleration
$\ve{a}_{A}$ of other bodies, iii) the mass $M_E$ of the Earth, iv) the
masses $M_A$ of other bodies, v) the higher-order multiple moments
$M_{L}$, $l\ge 2$ of the Earth. Note that in Newtonian physics
only the positions of the Earth and body $A$ ($\ve{x}_E$ and $\ve{x}_A$)
appear in $g_L^A$. In the post-Newtonian approximation one has

\begin{eqnarray}\label{qL}
g_{L}^A=&&\left.\left.\left.{(-1)^l\,(2l-1)!!\over r_{EA}^{l+1}}\,
\right[\,
\hat n_{EA}^L\,\right\{1+{1\over c^2}
\right(\,2\,v_{EA}^2-{1\over 2}\,\ve{a}_A\,\ve{r}_{EA}
-l\,\overline{w}_E(\ve{x}_E)-\overline{w}_A(\ve{x}_A)
\nonumber\\
&&
\left.\left.
\phantom{{(-1)^l\,(2l-1)!!\over r_{EA}^{l+1}}\,\Biggl[\,\hat n_{EA}^L\,\Biggl\{1+{1\over c^2}\Biggl(\,}
-{1\over 2}\,(2l+1)\,{\left(\ve{v}_A\,\ve{n}_{EA}\right)}^2
\right)\right\}
\nonumber\\
&&
\qquad
-{1\over c^2}\,{(l-1)\,(l-8)\over 2\,(2l-1)}\ \
v_{EA}^{\langle i_l}\,v_{EA}^{i_{l-1}}\,n_{EA}^{L-2\rangle}
+{1\over c^2}\,{1\over 2l-1}\,
r_{EA}\,a^{\langle i_l}\,n_{EA}^{L-1\rangle}
\nonumber\\
&&
\qquad
+{1\over c^2}\,{l\over 2}\
(\ve{v}_{E}\,\ve{n}_{EA})\,v_E^{\langle i_l}\,n_{EA}^{L-1\rangle}
\left.
-{1\over c^2}\
\left(l\,\ve{v}_{A}\,\ve{n}_{EA}+4\ve{v}_{EA}\,\ve{n}_{EA}\right)\,
v_{EA}^{\langle i_l}\,n_{EA}^{L-1\rangle}
\right],
\end{eqnarray}

\noindent
where
\begin{eqnarray}\label{qL-a}
&&\ve{a}=\left(l^2-l+4\right)\,\ve{a}_E+{1\over2}\,\left(l-8\right)\,\ve{a}_A,
\\
\label{qL-w-E-ext}
&&\overline{w}_E(\ve{x}_E)=\sum_{B\neq E}\,{GM_B\over r_{EB}},
\\
\label{qL-w-A-ext}
&&\overline{w}_A(\ve{x}_A)=\sum_{B\neq A}\,{GM_B\over r_{AB}}
+G\,\sum_{l=2}^\infty {(-1)^l\,(2l-1)!!\over l!\,r_{EA}^{l+1}}\,M_L\,\hat n_{EA}^L,
\end{eqnarray}

\noindent
and for any $A$ and $B$ one has $\ve{r}_{AB}=\ve{x}_A-\ve{x}_B$,
$\ve{v}_{AB}=\ve{v}_A-\ve{v}_B$,
$n^L_{AB}=\displaystyle{r_{AB}^{a_1}\dots r_{AB}^{a_l}\over r_{AB}^l}$.

\end{itemize}

\vspace*{1cm}

\noi {\large 6. Reduced equations of motion}

\vspace*{5mm}

Taking into account all the components of the model and the simplifying
assumptions the equation of rotational motion of the Earth with respect to
the GCRS can be written as

\begin{eqnarray}\label{reduced-eqm}
{d\over d\,\TCG}\,\left(C^{ab}\,\omega^b\right)=
\sum_{l=1}^\infty\ {1 \over l!}\
\varepsilon_{abc}\, M_{bL}\, G_{cL}
+\varepsilon_{abc}\,\Omega^b_{\rm iner}\,C^{cd}\,\omega^d.
\end{eqnarray}

The GCRS is kinematically non-rotating and this is the reason why
the post-Newtonian Coriolis force proportional to $\vecg{\Omega}_{\rm iner}$
appears in the right-hand side of
(\ref{reduced-eqm}). The equations of rotational motion of the Earth
relative to a dynamically non-rotating local geocentric reference
system does not contain this additional torque. However, the use of a
dynamically non-rotating reference system does not seem to be
advantageous since the slow precession of its spatial axes relative to
those of the BCRS must be taken into account
while computing the external tidal
moments $G_L$ in the dynamically non-rotating coordinates,
which is by no means simpler than using
(\ref{reduced-eqm}). Note also that the relative orientation of the
GCRS and that local dynamically non-rotating reference system is
well known (see, e.g., Brumberg, Bretagnon, Francou (1991))
and this can be used as a check of
theories of precession and nutation constructed in these two reference
systems. However, it does not mean that a theory of precession and
nutation of the Earth in one of these two reference systems can be
constructed in a purely Newtonian way, as it was assumed in
all modern theories of Earth nutation, where a purely Newtonian
theory was interpreted as a theory in dynamically non-rotating
coordinates.

The quantities characterizing the Earth, $C^{ab}$, $\omega^a$ and
$M_L$, are functions of \TCG\ or \TT\ while the quantities appearing in
the external tidal moments $G_L$ (e.g., the BCRS positions of the
bodies) are functions of \TCB\ or \TDB\ (since BCRS Solar system
ephemerides should be used here to evaluate those quantities). To avoid
the recomputing of $G_L$ from \TCB\ (or \TDB) to \TCG\ (or \TT) during
the [numerical] integration one may want to use \TDB\ as the
independent variable in (\ref{reduced-eqm}). This version of the
equations of rotational motion reads

\begin{eqnarray}\label{reduced-eqm-TDB}
{d\over d\,\TDB}\,\left(C^{ab}\,\omega^b\right)=
\left.\left({d\,\TCG\over d\,\TDB}\right)\right|_{\rm\ geocenter}
\,\cdot\,
\sum_{l=1}^\infty\ {1 \over l!}\
\varepsilon_{abc}\, M_{bL}\, G_{cL}
+\varepsilon_{abc}\,\Omega^b_{\rm iner}\,C^{cd}\,\omega^d.
\end{eqnarray}

A solution of this equation gives the Euler angles parametrizing the
orthogonal matrix $P^{ab}$ from (\ref{ovCij})--(\ref{ovM}) as functions
of \TDB\ which should be re-calculated as functions of \TCG\ or \TT\
afterwards.

The factor ${d\,\TCG\over d\,\TDB}$
gives a scaling of the torque as well as additional periodic signal in the torque.
Its practical importance should be further investigated.

Another important issue is that the equations (\ref{reduced-eqm}) and
the formulas  (\ref{qL})--(\ref{qL-w-A-ext}) for $G_L$ are valid only
if both time and space coordinates are not scaled in both GCRS and
BCRS. Since in practice one employs scaled time scales \TT\ and \TDB\
as well as associated scaled spatial coordinates, the corresponding
scaling factors must be taken into account while computing $G_L$ and
using (\ref{reduced-eqm}) or (\ref{reduced-eqm-TDB}).

In a further publication the equations of rotational motion
(\ref{reduced-eqm-TDB}) will be re-written in a form which can be
directly used for practical construction of a post-Newtonian theory of
Earth's rotation along the lines of Bretagnon {\it et al.} (1997,
1998).

\vspace*{1cm}

\noi {\large 7. REFERENCES} \\
{\leftskip=5mm
\parindent=-5mm

\ssk

%\noi
Bretagnon, P., Francou, G., Rocher, P., Simon, J.L. (1997)
Theory of the rotation of the rigid Earth
{\it Astronomy \& Astrophysics}, {\bf 319}, 305--317

%\noi
Bretagnon, P., Francou, G., Rocher, P., Simon, J.L. (1998)
SMART97: a new solution for the rotation of the rigid Earth
{\it Astronomy \& Astrophysics}, {\bf 329}, 329--338

%\noi
Brumberg, V.A., Bretagnon, P., Francou, G. (1991) Analytical algorithms
of relativistic reduction of astronomical observations.  In: {\sl
M\'etrologie et Astrom\'etrie}, edited by Capitaine, N., Proceedings of
Journ\'ees'1991, Observatoire de Paris, Paris, 141--148

%\noi
Damour, T., Soffel, M., Xu, C. (1993)
General Relativistic Celestial Mechanics III. Rotation Equations of Motion
{\it Phys. Rev. D}, {\bf Vol.~no.~47}, 3124--3135

%\noi
Fock, V.A. (1955)
{\sl Teoria Prostranstva, Vremeni i Tyagoteniya}, Fizmatgiz, Moscow
(translated into English as {\sl Theory of space, time and gravitation},
Pergamon, Oxford, 1959)

%\noi
IAU (2001) Information Bulletin No. 88, January 2001 (see also Erratum
in Information Bulletin No. 89, June 2001 and the full text of the
Resolution at \hfill\break
{\tt http://danof.obspm.fr/IAU\underline{\ }resolutions/Resol-UAI.htm}

%\noi
Klioner S.A. (1996):
Angular velocity of extended bodies in general relativity.
In: S.Ferraz-Mello, B.Morando, J.E.Arlot (eds.), Dynamics, ephemerides
and astrometry in the solar system, Kluwer, Dordrecht, 309--320

%\noi
Soffel, M. (1994)
The problem of rotational motion and rigid bodies in the post-Newtonian
framework.
{\it unpublished notes}

%\noi
Thorne, K.S., G\"ursel, Y. (1983)
The free precession of slowly rotating neutron stars: rigid-body motion
in general relativity,
{\it Mon. Not.  R astr. Soc.}, {\bf 205}, 809--817

%\noi
Voinov, A.V. (1988)
Motion and rotation of celestial bodies in the post-Newtonian approximation,
{\it Celestial Mechanics}, {\bf 41}, 293--307

}

\end{document}